\documentclass[5p]{elsarticle}
\biboptions{sort&compress}

\usepackage{amsmath}
\usepackage{upgreek}

\journal{JCG}

\begin{document}

\begin{frontmatter}


\title{Phase selection and microstructure of slowly solidified Al-Cu-Fe alloys}


\author[1,2]{L.V.~Kamaeva\corref{cor1}}

\author[1]{I.V.~Sterkhova}

\author[1]{V.I.~Lad`yanov}

\author[3,4,2]{R.E.~Ryltsev}

\author[2,4,5]{N.M.~Chtchelkatchev}

\address [1]{Udmurt Federal Research Center, Ural Branch of Russian Academy of Sciences, 426068 Izhevsk, Russia}
\address [2]{Vereshchagin Institute for High Pressure Physics, Russian Academy of Sciences, 108840 Troitsk, Moscow, Russia}
\address [3]{Institute of Metallurgy, Ural Branch of Russian Academy of Sciences, 620016, Ekaterinburg, Russia}
\address [4]{Ural Federal University, 620002, Ekaterinburg, Russia}
\address [5]{Moscow Institute of Physics and Technology, 141700, Dolgoprudny, Moscow Region, Russia}

\cortext[cor1]{Corresponding author}

\begin{abstract}
The search for effective methods to fabricate bulk single-phase quasicrystalline Al-Cu-Fe alloys is currently an important task. Crucial to solving this problem is to understand mechanisms of phase formation in this system. Here we study crystallization sequence during solidification as well as the conditions of solid phase formation in slowly solidified  Al-Cu-Fe alloys in a wide range of compositions. Concentration dependencies of undercoolability were also constructed by differential thermal analysis method. These experimental results are compared with data on chemical short-range order in the liquid state determined from \textit{ab initio} molecular dynamic simulations. We observe that main features of interatomic interaction in the Al-Cu-Fe alloys are similar for both liquid and solid states and they change in the vicinity of i-phase composition. In the concentration region, where the i-phase forms from the melt, both the undercoolability and the crystallization character depend on the temperature of the melts before cooling.
\end{abstract}

\end{frontmatter}

\section{Introduction}

The ${\rm Al-Cu-Fe}$  alloys are attractive model objects for studying the formation of a quasicrystalline state and conditions for its existence. The ${\rm Al-Cu-Fe}$ system is one of the first metallic systems in which the formation of stable icosahedral phase (i-phase) was reported~\cite{Tsai}. The icosahedral quasicrystals (IQCs) demonstrate unique properties and so investigation of their nucleation and growth processes is of special interest~\cite{Guiping, Gille, Baizuweit}. The Al-Cu-Fe i-phase can be obtained by rapid quenching of a melt (for example, by spinning or sputtering technique), mechanical alloying and plasma deposition methods~\cite{H}. These methods produce IQC in the form of either powders or thin coatings those applicability is essentially restricted. Therefore, the search for methods to fabricate bulk single-phase quasicrystalline Al-Cu-Fe alloys is currently an importent task. A promising way to do that is the use of structural heredity between liquid and solid states~\cite{Tasci, Brazhkin, Cai}. The general assumption is that the unusual structure of Al-based melts, including ${\rm Al-Cu-Fe}$ ones,  contributes to the formation of complex crystalline and quasi-crystalline phases~\cite{ Wang, Debela,Holland-Moritz}.
One of the main experimentally measured quantity, which is directly related to structural heredity, is undercoolability ($\Delta T$). On the one hand, $\Delta T$ depends on the initial structure of a liquid phase, and on the other hand, it controls concurrent processes of nucleation and growth of a solid phase. The impact of undercoolability  on the selection of phase formation in the ${\rm Al-Cu-Fe}$  system  has been insufficiently studied; only data for Al$_{62}$Cu$_{25.5}$Fe$_{12.5}$ and Al$_{60}$Cu$_{34}$Fe$_{6}$ alloys have been reported so far~\cite{Holland-Moritz}.

Here we address this issue for slowly solidified  Al-Cu-Fe alloys in a wide range of compositions. We consider two sections  Al$_{57+x}$Cu$_{40.5-x}$Fe$_{12.5}$ and Al$_{52+x}$Cu$_{25.5}$Fe$_{22.5-x}$, ($x=0-20$ at.\%), containing i-phase stoichiometry composition, as well as concentration range Al$_{58.7+x}$Cu$_{35.5-x}$Fe$_{5.8}$, ($x=0-15$) where the i-phase nucleates from the melt. For these alloys, we study crystallization sequence during solidification as well as the conditions of solid phase formation. The results obtained are compared with \textit{ab initio} molecular dynamic simulations data for the structure of melts at different concentrations. It allows us to establish the relationship between practically important solidification processes in ${\rm Al-Cu-Fe}$  alloys and the structure of their melts.

\section{Material and methods}

The samples under study were prepared by alloying the Al-Fe ligatures (Al$_{80.6}$Fe$_{19.4}$ or Al$_{69.8}$Fe$_{30.2}$ depending on the alloy composition), A999-type Al and Cu cathode  in a vacuum furnace under a helium atmosphere (after preliminary pumping to 10$^{-3}$ Pa) at 1400 $^\circ$C for 1 hour.

The undercoolability values were determined by the Differential Thermal Analysis (DTA) technique. The DTA experiments were carried out using a high temperature thermal analyzer which comprises a chamber (a high temperature vacuum resistance furnace with a Mo heater and a thermostat located in the furnace), the chamber evacuation and the gas release equipments. The thermostat contains two cells equipped with the Al$_2$O$_3$ crucibles filled with either a standard material  (W) or a sample under study. The temperature difference $dT_{DTA}$ between the standard sample and the studied one is a measured parameter of the method. The experiments were performed in pure helium conditions under a low excess pressure after evacuation down to a pressure of about 10$^{-2}$ Pa.

Metallography was carried out using optical microscope after the standard surface preparation procedure, namely chemical etching in various reagents depending on the alloy compositions (R1 - 3.5 g FeCl$_{3}$, 25 ml HCl, 75 ml C${2}$H$_{5}$OH; R2 - 10 ml H$_{2}$O, 1 g FeSO$_{4}$; R3 - 95 ml H$_{2}$O, 2.5 ml HNO$_{3}$, 1.5 ml HCl, 1ml HF). Qualitative structural analysis of the samples was performed by X-ray diffraction (XRD) method using Fe radiation.

The \textit{ab initio} molecular dynamics (AIMD) simulations were performed based on the density functional theory as implemented in the CP2K package \cite{MD1}.  Projector augmented-wave (PAW) pseudopotentials and Perdew-Burke-Ernzerhof gradient approximation to the exchange-correlation functional were used \cite{MD2}.  Cubic supercell of 512 atoms with the composition Al$_{68.7}$ Cu$_{25.5}$Fe$_{5.8}$, Al$_{69.5}$Cu$_{18}$Fe$_{12.5}$, Al$_{62}$Cu$_{25.5}$Fe$_{12.5}$, Al$_{52}$Cu$_{35.5}$ Fe$_{12.5}$, Al$_{57}$Cu$_{25.5}$Fe$_{17.5}$, Al$_{52}$Cu$_{25.5}$Fe$_{22.5}$ were build. Simulations were performed in the NVT ensemble at the density corresponding to  pressure $P = 0$, which was obtained by minimizing energy as a function of volume. The time step was 1 fs. The initial configuration of the melt was heated up to 10000K and balanced by performing several thousand MD steps, then relaxed at the desired temperature for several thousand MD steps.

\section{Results and discussion}

The DTA plots (thermograms) were obtained on heating up to 1400 $^\circ$ at 20 $^\circ$C/min. Then the sample was held for 20 min at this temperature and cooled. On the heating and cooling thermograms of Al-Cu-Fe alloys, there are from 3 to 6 endothermic and exothermic peaks, respectively. The temperatures of each step of melting (on heating) and crystallization (on cooling) for all studied alloys were determined using DTA thermograms. Fig.~1 (a, b)  shows the concentration dependence of melting temperature (liquidus) $T_L(x)$ the alloys investigated. These temperatures are in good agreement with the ternary state diagram available in \cite{diagrAlCuFe} and refine the calculated liquidus temperatures. Locations of minima on $T_L(x)$ for concentration cross sections of 12.5 at.\% Fe and 25.5 at.\% Cu coincide with i-phase concentration and correspond to invariant equilibrium reaction: L$\to$ L + Al$_{3}$Fe + $\beta$, where $\beta$ is either B2 or the disordered form (bcc).

\begin{figure}
\resizebox{1\columnwidth}{!}{%
 \includegraphics{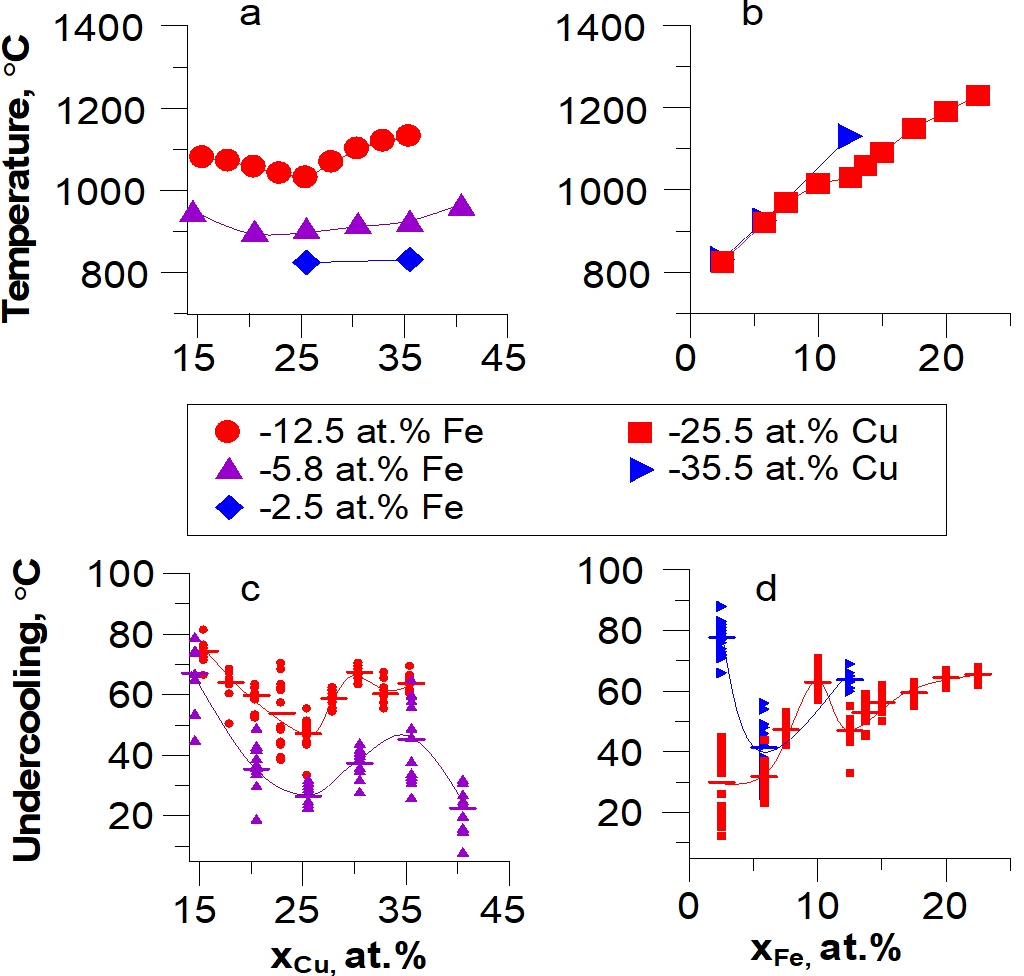} }
\caption{Concentration dependences of melting points (liquidus) of Al-Cu-Fe alloys (a, b) determined by heating DTA plot at 20  $^\circ$C/min and concentration dependences of undercoolingability under different cooling conditions of Al-Cu-Fe alloys (c, d): cooling from 1400$^\circ$C at 20, 100 and 50 $^\circ$C/min and during thermal cycling of the samples.}
\label{fig:1}       
\end{figure}

The undercoolability was calculated as the difference between the melting temperature and the crystallization temperature determined from heating and cooling thermograms, respectivelly. In the process of melt crystallization in a container, the undercoolability value is influenced by different factors, such as cooling rate, the temperature of the molten alloys and the number of melting–crystallization cycles \cite{DT}. Therefore, the experimental measurement process for each sample consisted of several heating (melting)-cooling (crystallization) cycles in which either the maximum heating temperature of melts ($T_{max}$) or the cooling rate was changed. The experimental cooling rates were 20, 50 and 100 $^\circ$C/min.  The impact of $T_{max}$ on undercoolability was studied in special experiments, thermal cycling, when the sample was heated to the temperature 15–25 $^\circ$C higher than the melt temperature, then it was held for 20 min and cooled at 100 $^\circ$C/min. In further heating–cooling cycle, the maximum heating temperature was increased at 10–20  $^\circ$C  higher than the previous and so on until the T$_{max}$ reached to 1400 $^\circ$C. The concentration dependencies of undercoolability ($\Delta T(x)$) obtained at cooling from different melt temperatures at various rates are presented in Fig. 1 (c, d). As seen from Fig. 1 (c, d) the melts are crystallized under low undercooling conditions (from 20 to 100 $^\circ$C). The $\Delta T(x)$ dependencies  for the studied concentration cross sections are in close correlation with concentration variation of liquidus temperature. The largest differences between these quantities are observed in the region with a low Fe content where peritectic reaction (L+ Al$_{3}$Fe + $\beta$$\to$i) takes place,  and and large Cu content where i-phase nucleates from the melt \cite{diagrAlCuFe}.

To determine the solid phases formation sequence during cooling, a metallographic analysis of the samples after DTA was carried out. The samples obtained demonstrate complex multiphase structures, which depend on the composition (Fig. 2). They are characterized by a large number of macrodefects (pores and cracks). This is caused by the formation of a quasicrystalline phase whose structure is disproportionate to the structure of other crystalline phases. The microstructures of the alloys are mainly formed as a result of peritectic reactions, which proceed one after the other.

\begin{figure}
\resizebox{1\columnwidth}{!}{%
 \includegraphics{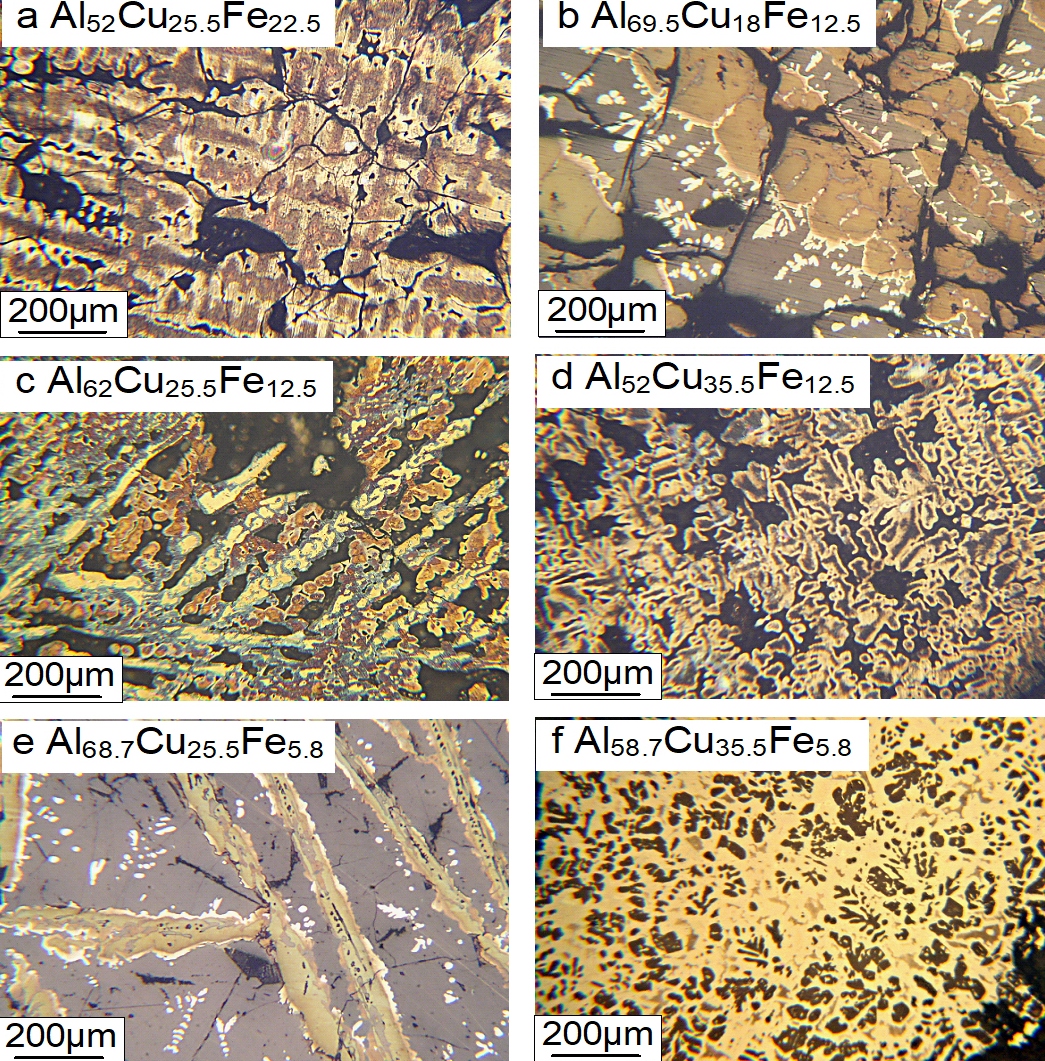} }
\caption{The microstructures of Al-Cu-Fe alloys after cooling from 1400 $^\circ$ at a rate of 100$^\circ$C/min. Microstructures were produced by etching in the R1 (a, d), R2 (b, c, e) and R3 (f).}
\label{fig:2}       
\end{figure}

The  ${\rm Al-Cu-Fe}$  alloys at 12.5 at.\% Fe with a copper content of up to 25.5 at.\% begin to solidify  with the formation of plate-like Al$_{3}$Fe crystals (fig. 2b). At the concentration range from 25.5 to 35 at.\% Cu, the rounded dendrites of the  $\beta$ solid solution grow in the first stage of solidification (fig. 2d). At 25.5 at.\% Cu, under conditions of minimal undercooling,  $\beta$ rounded dendrites and Al$_{3}$Fe lamellar crystals are formed from the melt, then the i-phase grows along their interfaces (fig. 2c). According to XRD, the ternary $\omega$ phase \cite{diagrAlCuFe} as well as Al-based solid solution are also presented in the composition of these alloys. The latter is presented in the Al-rich concentration region. Similar changes in both microstructure and the conditions of nucleation and growth of the Al$_{3}$Fe and $\beta$ phases are observed at 25. 5 at. \% Cu as iron concentration varies from 2.5 to 22 at.\% (fig. 2 a, c, e).
At 5.8 at.\% Fe, the microstructure of the alloys qualitatively changes if Cu concentrations is higher than 30 at.\%. Note that, at this concentration region, we observe the noticeable impact of the maximum heating temperature on the value of the undercoolability. This leads to an increase of the error of undercooling values (fig.1 c). The impact of the $T_{max}$ on undercoolability is illustrated in Fig. 3 by the example of Al$_{58.7}$Cu$_{35.5}$Fe$_{5.8}$ alloy.

\begin{figure}
\resizebox{1\columnwidth}{!}{%
 \includegraphics{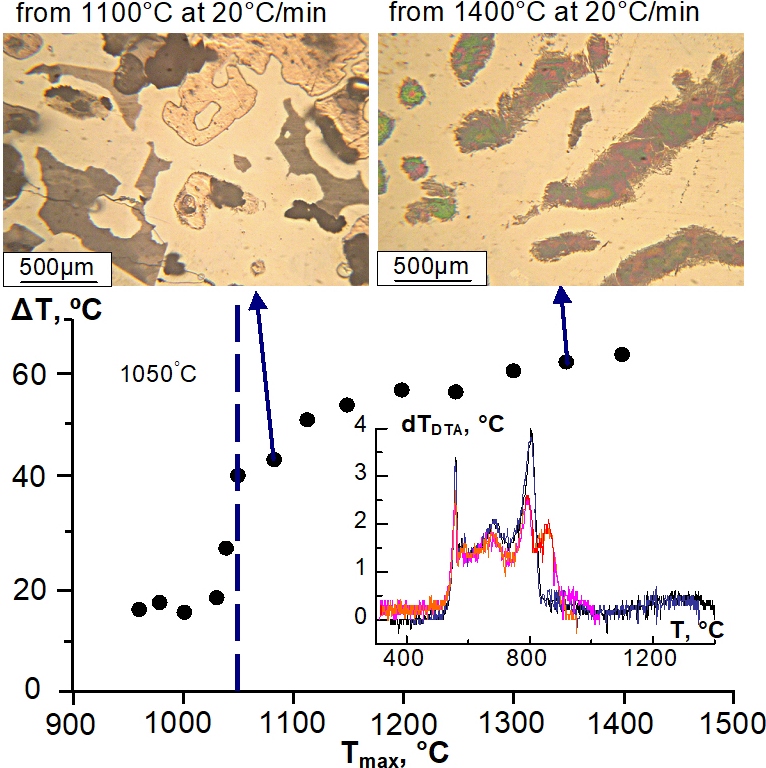} }
\caption{The microstructures, the cooling DTA plots and temperature dependence of undecooling ability of Al$_{58.7}$Cu$_{35.5}$Fe$_{5.8}$ alloy by cooling from different T$_{max}$ at 20$^\circ$C/min .}
\label{fig:3}       
\end{figure}

 It can be seen from the Fig. 3 that the type of cooling thermograms as well as the values of undercoolability depend on the temperature of the melt before cooling. When the melt is cooled from temperatures $T<$1050 $^\circ$C, the alloys  demonstrate low undercoolability and solidifies according to the equilibrium state diagram. With an increase of T$_{max}$ the undercoolability increases (more then 60 $^\circ$C), and the character of solidification changes so that first solidification stage corresponding to the formation of the i-phase is absent on both thermograms and microstructure images (fig. 3).

To analyze structural changes in Al-Cu-Fe melts under change of composition, we perform \textit{ab initio} MD  simulations at seven different concentrations.  The temperature for each alloy is about 100 K above its melting points. According to calculated total and partial coordination numbers, the Warren–Cowley chemical short-range order (CSRO) parameter  $\alpha$$_{i–j}$ \cite{Debela} is calculated to describe the chemical environment around the central atom $i$ ($i$ = Al, Cu). The parameter is defined as $$\alpha_{i–j}=1-\frac{Z_{i-j}}{Z_{i-total}x_{j}}$$ where $Z_{i-total}$ and $Z_{i–j}$ are respectively the total and partial coordination numbers, $x_{j}$ is the concentration of coordination atom $j$. Two groups of $\alpha$$_{i–j}$, classified by the type of central atoms, are plotted in Fig. 4. For random distribution, $\alpha$$_{i–j}$ is equal to zero. The negative $\alpha$$_{i–j}$ means the attraction between $i$ and $j$, while the positive $\alpha$$_{i–j}$ reflects repulsion.

As shown in Fig. 4, the strongest chemical interaction in the melts studied is the repulsion between Cu and Fe atoms, that is in good agreement with the data on the short-range order in solid Al-Cu-Fe alloys \cite{B2}. The concentration changes of the CSRO around Cu atoms shows that the repulsion between Cu and Fe atoms has minimal values at concentrations of iron and copper corresponding to the i-phase stoichiometry. It is in good agreement with the concentration behavior of the melting temperature and undercoolability $\Delta T(x)$.  The increase of copper concentration at 12.5 at.\%Fe and iron at 25.5 at.\% Cu in Al surroundings leads to a change of the interaction type, both attraction between Al atoms and repulsion between Al and Cu, Al, Fe atoms decreases  and change their sign in the i-phase stoichiometry region (this is clearly seen for Fe atoms in Fig.~4). At high concentrations of both Cu and Fe, we see almost random distribution around Al atoms without the CSRO. The intensive chemical interaction is observed in the region of low Cu and Fe concentrations where our experiment show essential dependence of underrcoolingability and phase formation order on the temperature of the melts.

\section{Conclusions}

\begin{figure}
\resizebox{1\columnwidth}{!}{%
 \includegraphics{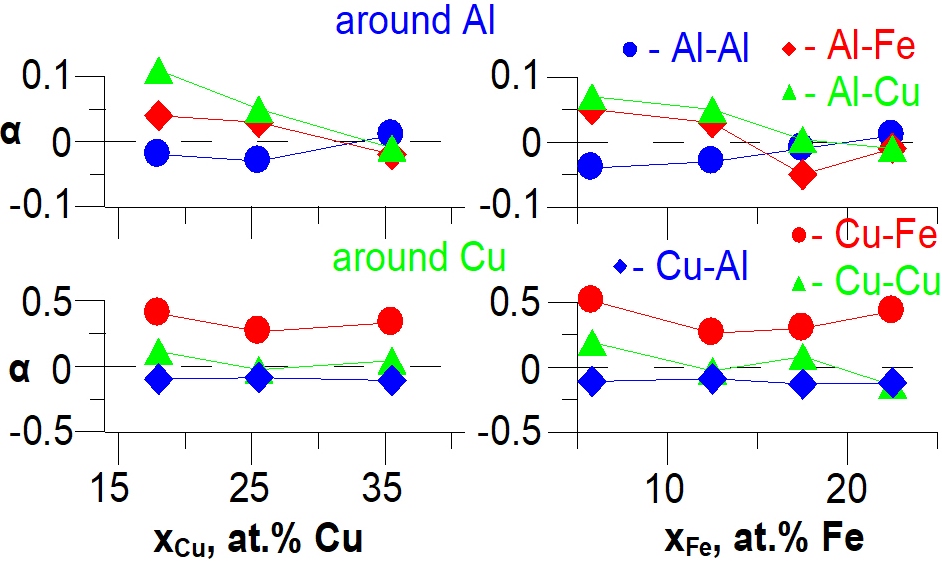} }
\caption{The concentration behavior of the coefficients of the chemical short-range order $\alpha$ (the parameter  Warren–Cowley) according MD.}
\label{fig:4}       
\end{figure}

Analysis of the CSRO reveals that main features of interatomic interaction in the Al-Cu-Fe alloys are similar for both liquid and solid states. In the vicinity of Cu and Fe concentrations corresponding to the composition of the i-phase, a change in CSRO is observed. This effect is consistent with the minimum on the liquidus line and affects essentially the initial stage of solidification.

In the concentration region, where the i-phase forms from the melt, both the  undercoolability and the crystallization character depend on the temperature of the melts before cooling. At cooling of the melt from temperatures less than 1050-1100 $^\circ$C, the alloys solidify under low undercooling conditions according to the equilibrium state diagram. An increase of melt temperature leads to an increase in undercoolability (more than 60 $^\circ$C) and suppress formation of the i-phase.

\section{Acknowledgments}
This work was supported by Russian Science Foundation (grant RNF 18-12-00438). AIMD simulations have been carried out using  "Uran"\ supercomputer of IMM UB RAS and computing resources of the federal collective usage center Complex for Simulation and Data Processing for Mega-science Facilities at NRC “Kurchatov Institute” {http://ckp.nrcki.ru}

\bibliography{mybibfile1}

\begin{thebibliography}{16}
\expandafter\ifx\csname natexlab\endcsname\relax\def\natexlab#1{#1}\fi
\providecommand{\url}[1]{\texttt{#1}}
\providecommand{\href}[2]{#2}
\providecommand{\path}[1]{#1}
\providecommand{\DOIprefix}{doi:}
\providecommand{\ArXivprefix}{arXiv:}
\providecommand{\URLprefix}{URL: }
\providecommand{\Pubmedprefix}{pmid:}
\providecommand{\doi}[1]{\href{http://dx.doi.org/#1}{\path{#1}}}
\providecommand{\Pubmed}[1]{\href{pmid:#1}{\path{#1}}}
\providecommand{\bibinfo}[2]{#2}
\ifx\xfnm\relax \def\xfnm[#1]{\unskip,\space#1}\fi
\bibitem[{Tsai et~al.(1987)Tsai, Inoue, and Masumoto}]{Tsai}
\bibinfo{author}{A.-P. Tsai}, \bibinfo{author}{A.~Inoue},
  \bibinfo{author}{T.~Masumoto},
\newblock \bibinfo{title}{{A Stable Quasicrystal in Al-Cu-Fe System}},
\newblock \bibinfo{journal}{Jpn. J. Appl. Phys.} \bibinfo{volume}{26}
  (\bibinfo{year}{1987}) \bibinfo{pages}{L1505}.
\bibitem[{Ji et~al.(2019)Ji, Ji, Lu, and Cao}]{Guiping}
\bibinfo{author}{G.~Ji}, \bibinfo{author}{A.~Ji}, \bibinfo{author}{N.~Lu},
  \bibinfo{author}{Z.~Cao},
\newblock \bibinfo{title}{{Formation and morphology evolution of icosahedral
  and decahedral silver crystallites from vapor deposition in view of symmetry
  misfit}},
\newblock \bibinfo{journal}{J. Cryst. Growth} \bibinfo{volume}{518}
  (\bibinfo{year}{2019}) \bibinfo{pages}{89}.
\bibitem[{Gille et~al.(2011)Gille, Bauer, Hahne, Smontara, and
  Dolinsek}]{Gille}
\bibinfo{author}{P.~Gille}, \bibinfo{author}{B.~Bauer},
  \bibinfo{author}{M.~Hahne}, \bibinfo{author}{A.~Smontara},
  \bibinfo{author}{J.~Dolinsek},
\newblock \bibinfo{title}{{Single
  crystalgrowthofAl-basedintermetallicphasesbeingapproximants to
  quasicrystals}},
\newblock \bibinfo{journal}{J. Cryst. Growth} \bibinfo{volume}{318}
  (\bibinfo{year}{2011}) \bibinfo{pages}{1016}.
\bibitem[{Baizuweit and Enckevort(1994)}]{Baizuweit}
\bibinfo{author}{K.~Baizuweit}, \bibinfo{author}{W.~J.~P. Enckevort},
\newblock \bibinfo{title}{{Surface studies on the five-fold facets of
  Al—Cu—Fe quasicrystals}},
\newblock \bibinfo{journal}{J. Cryst. Growth} \bibinfo{volume}{135}
  (\bibinfo{year}{1994}) \bibinfo{pages}{297}.
\bibitem[{Huttunen-Saarivirta(2004)}]{H}
\bibinfo{author}{E.~Huttunen-Saarivirta},
\newblock \bibinfo{title}{{Microstructure, fabrication and properties of
  quasicrystalline Al–Cu–Fe alloys}},
\newblock \bibinfo{journal}{J. Alloys Compd.} \bibinfo{volume}{363}
  (\bibinfo{year}{2004}) \bibinfo{pages}{150--174}.
\bibitem[{Tasci et~al.(2010)Tasci, Sluiter, Pasturel, and Villars}]{Tasci}
\bibinfo{author}{E.~S. Tasci}, \bibinfo{author}{M.~H.~F. Sluiter},
  \bibinfo{author}{A.~Pasturel}, \bibinfo{author}{P.~Villars},
\newblock \bibinfo{title}{{Liquid structure as a guide for phase stability in
  the solid state: Discovery of a stable compound in the Au–Si alloy
  system}},
\newblock \bibinfo{journal}{Acta Mater.} \bibinfo{volume}{58}
  (\bibinfo{year}{2010}) \bibinfo{pages}{449}.
\bibitem[{Brazhkin et~al.(2008)Brazhkin, Katayama, Kondrin, Yattori, Lyapin,
  and Saitoh}]{Brazhkin}
\bibinfo{author}{V.~V. Brazhkin}, \bibinfo{author}{Y.~Katayama},
  \bibinfo{author}{M.~V. Kondrin}, \bibinfo{author}{T.~Yattori},
  \bibinfo{author}{A.~G. Lyapin}, \bibinfo{author}{H.~Saitoh},
\newblock \bibinfo{title}{{AsS melt under pressure: one substance, three
  liquids}},
\newblock \bibinfo{journal}{Phys. Rev. Lett} \bibinfo{volume}{100}
  (\bibinfo{year}{2008}) \bibinfo{pages}{145701}.
\bibitem[{Cai et~al.(2001)Cai, Zhang, Li, Chen, and Fu}]{Cai}
\bibinfo{author}{Y.~Cai}, \bibinfo{author}{G.~Zhang}, \bibinfo{author}{J.~Li},
  \bibinfo{author}{G.~Chen}, \bibinfo{author}{H.~Fu},
\newblock \bibinfo{title}{{Modeling Elasticity in Crystal Growth}},
\newblock \bibinfo{journal}{Sci. Technol. Adv. Mater.} \bibinfo{volume}{2}
  (\bibinfo{year}{2001}) \bibinfo{pages}{169--172}.
\bibitem[{Wang et~al.(2017)Wang, Li, Li, Pan, and Qin}]{Wang}
\bibinfo{author}{J.~Wang}, \bibinfo{author}{X.~Li}, \bibinfo{author}{J.~Li},
  \bibinfo{author}{S.~Pan}, \bibinfo{author}{J.~Qin},
\newblock \bibinfo{title}{{Mg fragments and Al bonded networks in liquid
  Mg–Al alloys}},
\newblock \bibinfo{journal}{Comput. Mater. Sci.} \bibinfo{volume}{129}
  (\bibinfo{year}{2017}) \bibinfo{pages}{115–122}.
\bibitem[{Debela et~al.(2018)Debela, Abbas, Pasturel, and Villars}]{Debela}
\bibinfo{author}{T.~T. Debela}, \bibinfo{author}{H.~G. Abbas},
  \bibinfo{author}{A.~Pasturel}, \bibinfo{author}{P.~Villars},
\newblock \bibinfo{title}{{ Role of nanosize icosahedral quasicrystal of Mg-Al
  and Mg-Ca alloys in avoiding crystallization of liquid Mg: Ab initio
  molecular dynamics study}},
\newblock \bibinfo{journal}{J. Non-Cryst. Solids} \bibinfo{volume}{449}
  (\bibinfo{year}{2018}) \bibinfo{pages}{173--182}.
\bibitem[{Holland-Moritz et~al.(1997)Holland-Moritz, Schroers, Grushko,
  Herlach, and Urban}]{Holland-Moritz}
\bibinfo{author}{D.~Holland-Moritz}, \bibinfo{author}{J.~Schroers},
  \bibinfo{author}{B.~Grushko}, \bibinfo{author}{D.~M. Herlach},
  \bibinfo{author}{K.~Urban},
\newblock \bibinfo{title}{{Dependence of phase selection and microstructure of
  quasicrystal-forming Al-Cu-Fe alloys on the processing and solidification
  conditions}},
\newblock \bibinfo{journal}{J. Non-Cryst. Solids} \bibinfo{volume}{A226-228}
  (\bibinfo{year}{1997}) \bibinfo{pages}{976--980}.
\bibitem[{Qin et~al.(2011)Qin, Geng, and Li}]{MD1}
\bibinfo{author}{H.~O. Qin}, \bibinfo{author}{H.~R. Geng},
  \bibinfo{author}{Z.~Y. Li},
\newblock \bibinfo{title}{{in Applied Mechanics and Materials}},
\newblock \bibinfo{journal}{Trans. Tech. Pub.} \bibinfo{volume}{55}
  (\bibinfo{year}{2011}) \bibinfo{pages}{913--917}.
\bibitem[{Perdew et~al.(1996)Perdew, Burke, and Ernzerhof}]{MD2}
\bibinfo{author}{J.~P. Perdew}, \bibinfo{author}{K.~Burke},
  \bibinfo{author}{M.~Ernzerhof},
\newblock \bibinfo{title}{{Generalized Gradient Approximation Made Simple}},
\newblock \bibinfo{journal}{Phys. Rev. Lett.} \bibinfo{volume}{77}
  (\bibinfo{year}{1996}) \bibinfo{pages}{3865--3868}.
\bibitem[{Raghavan(2005)}]{diagrAlCuFe}
\bibinfo{author}{V.~Raghavan},
\newblock \bibinfo{title}{{Al-Cu-Fe (Aluminum-Copper-Iron)}},
\newblock \bibinfo{journal}{J. Phase Equilib. Diffus.} \bibinfo{volume}{26}
  (\bibinfo{year}{2005}) \bibinfo{pages}{59--64}.
\bibitem[{Zhou et~al.(2000)Zhou, Wang, and Sun}]{DT}
\bibinfo{author}{Z.~Zhou}, \bibinfo{author}{W.~Wang}, \bibinfo{author}{B.~Sun},
\newblock \bibinfo{title}{{Undercooling and metastable phase formation in a
  Bi95Sb5 melt}},
\newblock \bibinfo{journal}{Appl. Phys. A} \bibinfo{volume}{261-265}
  (\bibinfo{year}{2000}) \bibinfo{pages}{261--265}.
\bibitem[{Brand et~al.(2001)Brand, Voss, and Calvayrac}]{B2}
\bibinfo{author}{R.~A. Brand}, \bibinfo{author}{J.~Voss},
  \bibinfo{author}{Y.~Calvayrac},
\newblock \bibinfo{title}{{Dynamics in the icosahedral quasicrystal
  i-Al$_{62}$Cu$_{25.5}$Fe$_{12.5}$: phonons and phasons}},
\newblock \bibinfo{journal}{J. Non-Cryst. Solids} \bibinfo{volume}{287}
  (\bibinfo{year}{2001}) \bibinfo{pages}{210--215}.

\end{thebibliography}

\end{document}